\documentclass{emulateapj}

\usepackage{amsmath}

\newcommand{\about}{$\sim\!\!$~}

\newcommand{\be}{\begin{displaymath}}
\newcommand{\ee}{\end{displaymath}}

\def\lsim{\hbox{\rlap{\raise 0.425ex\hbox{$<$}}\lower 0.65ex\hbox{$\sim$}}}
\def\gsim{\hbox{\rlap{\raise 0.425ex\hbox{$>$}}\lower 0.65ex\hbox{$\sim$}}}
\def\arcmin{\hbox{$^\prime$}}
\def\arcsec{\hbox{$^{\prime\prime}$}}
\def\arcdeg{\mbox{$^\circ$}}

\newcommand{\msun}{M$_\sun$}

\newcommand{\kms}{km\,s$^{-1}$}

\shorttitle{The Young Type Ia SN 2016coj}
\shortauthors{Zheng et al.}

\begin{document}

\title{Discovery and Follow-up Observations of the Young Type Ia Supernova 2016coj}

\author{WeiKang Zheng\altaffilmark{1,2},        
Alexei V. Filippenko\altaffilmark{1},           
Jon Mauerhan\altaffilmark{1},                   
Melissa L. Graham\altaffilmark{1,3},               
Heechan Yuk\altaffilmark{1},                    
Griffin Hosseinzadeh\altaffilmark{4,5},         
Jeffrey M. Silverman\altaffilmark{6,7},         
Liming Rui\altaffilmark{8},                     
Ron Arbour\altaffilmark{9},                     
Ryan J. Foley\altaffilmark{10},                 
Bela Abolfathi\altaffilmark{11},                
Louis E. Abramson\altaffilmark{12},             
Iair Arcavi\altaffilmark{4,13},                 
Aaron J. Barth\altaffilmark{11},                
Vardha N. Bennert\altaffilmark{14},             
Andrew P. Brandel\altaffilmark{11},             
Michael C. Cooper\altaffilmark{11},             
Maren Cosens\altaffilmark{14},                  
Sean P. Fillingham\altaffilmark{11},            
Benjamin J. Fulton\altaffilmark{15},            
Goni Halevi\altaffilmark{1},                    
D. Andrew Howell\altaffilmark{4,5},             
Tiffany Hsyu\altaffilmark{10},                  
Patrick L. Kelly\altaffilmark{1},               
Sahana Kumar\altaffilmark{1},                   
Linyi Li\altaffilmark{8},                       
Wenxiong Li\altaffilmark{8},                    
Matthew A. Malkan\altaffilmark{12},             
Christina Manzano-King\altaffilmark{16},        
Curtis McCully\altaffilmark{4,5},               
Peter E. Nugent\altaffilmark{1,17},             
Yen-Chen Pan\altaffilmark{10},                  
Liuyi Pei\altaffilmark{11},                     
Bryan Scott\altaffilmark{16},                   
Remington Oliver Sexton\altaffilmark{16},       
Isaac Shivvers\altaffilmark{1},                 
Benjamin Stahl\altaffilmark{18},                
Tommaso Treu\altaffilmark{12},                  
Stefano Valenti\altaffilmark{19},               
H. Alexander Vogler\altaffilmark{11},          
Jonelle L. Walsh\altaffilmark{20},              
Xiaofeng Wang\altaffilmark{8}                   
}

\altaffiltext{1}{Department of Astronomy, University of California, Berkeley, CA 94720-3411, USA.}
\altaffiltext{2}{e-mail: zwk@astro.berkeley.edu .}
\altaffiltext{3}{Department of Astronomy, University of Washington, Box 351580, U.W., Seattle, WA 98195-1580, USA.}
\altaffiltext{4}{Las Cumbres Observatory Global Telescope Network, 6740 Cortona Dr Ste 102, Goleta, CA 93117-5575, USA.}
\altaffiltext{5}{Department of Physics, University of California, Santa Barbara, CA 93106-9530, USA.}
\altaffiltext{6}{Department of Astronomy, University of Texas, Austin, TX 78712, USA.}
\altaffiltext{7}{NSF Astronomy and Astrophysics Postdoctoral Fellow.}
\altaffiltext{8}{Physics Department and Tsinghua Center for Astrophysics, Tsinghua University, Beijing, 100084, China.}
\altaffiltext{9}{Pennel Observatory, 29 Wrights Way, South Wonston, Hants S021 3He}
\altaffiltext{10}{Department of Astronomy and Astrophysics, University of California, Santa Cruz, CA 95064, USA.}
\altaffiltext{11}{Department of Physics and Astronomy, University of California, 4129 Frederick Reines Hall, Irvine, CA 92697, USA.}
\altaffiltext{12}{Department of Physics and Astronomy, University of California, 430 Portola Plaza, Los Angeles, CA 90095, USA.}
\altaffiltext{13}{Kavli Institute for Theoretical Physics, University of California, Santa Barbara, CA 93106-4030, USA.}
\altaffiltext{14}{Physics Department, California Polytechnic State University, San Luis Obispo, CA 93407, USA.}
\altaffiltext{15}{Institute for Astronomy, University of Hawaii, 2680 Woodlawn Dr., Honolulu, HI 96822, USA.}
\altaffiltext{16}{Department of Physics and Astronomy, University of California, 900 University Avenue, Riverside, CA 92521, USA.}
\altaffiltext{17}{Lawrence Berkeley National Laboratory, Berkeley, California 94720, USA.}
\altaffiltext{18}{Department of Physics, University of California, Berkeley, 94720, USA.}
\altaffiltext{19}{Department of Physics, University of California, Davis, 1 Shields Ave, Davis, CA 95616-5270, USA.}
\altaffiltext{20}{George P. and Cynthia Woods Mitchell Institute for Fundamental Physics and Astronomy, Department of Physics and Astronomy, Texas A\&M University, College Station, TX 77843, USA.}

\begin{abstract}
The Type~Ia supernova (SN~Ia) 2016coj in NGC~4125 (redshift $z=0.004523$)
was discovered by the Lick Observatory Supernova Search 4.9 days after
the fitted first-light time (FFLT; 11.1 days before $B$-band maximum).
Our first detection (pre-discovery) is merely $0.6\pm0.5$ day after the FFLT,
making SN~2016coj one of the earliest known detections of a SN~Ia.
A spectrum was taken only 3.7 hr after discovery (5.0 days after the FFLT)
and classified as a
normal SN~Ia. We performed high-quality photometry, low- and
high-resolution spectroscopy, and spectropolarimetry,
finding that SN~2016coj is a spectroscopically normal SN~Ia,
but with a high velocity of \ion{Si}{2} $\lambda$6355 ($\sim 12,600$\,\kms\ around peak brightness).
The \ion{Si}{2} $\lambda$6355 velocity evolution can be well fit by a
broken-power-law function for up to a month after the FFLT.
SN~2016coj has a normal peak luminosity ($M_B \approx -18.9 \pm 0.2$ mag),
and it reaches a $B$-band maximum \about16.0~d after the FFLT.
We estimate there to be low host-galaxy extinction based on the
absence of Na~I~D absorption lines in our low- and high-resolution spectra.
The spectropolarimetric data exhibit weak polarization in the continuum,
but the \ion{Si}{2} line polarization is quite strong ($\sim 0.9\% \pm 
0.1\%$) at peak brightness.
\end{abstract}



\section{Introduction}\label{s:intro}

Type~Ia supernovae (SNe~Ia; see Filippenko 1997 for a review of supernova 
classification)
are the thermonuclear runaway explosions of carbon/oxygen white dwarfs
(see, e.g., Hillebrandt \& Niemeyer 2000 for a review).
They can be used as standardizable candles with
many important applications, including measurements of the expansion rate of the Universe
(Riess et al. 1998; Perlmutter et al. 1999). 
Two general scenarios are favored as the progenitor system for SNe~Ia. One
is the single-degenrate model (Hoyle \& Fowler 1960; Hachisu et al. 1996;
Meng et al. 2009; R\"{o}pke et al. 2012), which consists of a single white dwarf
accreting material from a companion. The other is the double-degenerate
scenario involving the merger of two white dwarfs (Webbink 1984;
Iben \& Tutukov 1984; Pakmor et al. 2012; R\"{o}pke et al. 2012).
However, our understanding of their progenitor systems and explosion mechanisms
remains substantially incomplete. 

Very early discovery and detailed follow-up observations
are essential for understanding those problems.
For example, Bloom et al. (2012) were able to constrain the companion-star radius
to be $\lesssim 0.1\,{\rm R}_\sun$ from an optical nondetection just
4\,hr after the explosion of SN~2011fe (Nugent et al. 2011).
Cao et al. (2015) found strong but declining ultraviolet emission in SN~Ia
iPTF14atg in early-time {\it Swift} observations,
consistent with theoretical expectations of the collision between supernova
(SN) ejecta and a companion star (Kasen 2010).
Im et al. (2015) found evidence of a ``dark phase" in SN~2015F, which can
last for a few hours to days between the moment of explosion and the first
observed light (e.g., Rabinak, Livne, \& Waxman 2012;
Piro \& Nakar 2013, 2014); see also Cao et al. (2016) for the case
of iPTF14pdk.
Differences in the duration of the ``dark phase" could be 
caused by a varying distribution of $^{56}$Ni
near the surface of a SN~Ia. For example, Piro \& Morozova (2016) show that
it is short (with a steep rise) when the $^{56}$Ni is shallow, and
longer (with a more gradual rise) when the $^{56}$Ni is deeper 
(see their Figure 7).

Spectra of SNe~Ia not only reveal the ejecta
composition from nuclear burning, but also provide a way to measure
the ejecta expansion velocity.
Benetti et al. (2005) separated SN~Ia samples into different groups
according to their velocity gradient and found that high-velocity-gradient
objects tend to have a higher velocity of the \ion{Si}{2} $\lambda$6355
line near maximum light.
Nugent et al. (1995) quantified the spectral diversity using line-strength
ratios, finding a good correlation
between the absorption-depth ratio of \ion{Si}{2} $\lambda$5972 to
\ion{Si}{2} $\lambda$6355) and the brightness decline rate.
Wang et al. (2009; 2013) and Foley et al. (2011a) separated SNe~Ia 
into high-velocity and normal-velocity groups with a boundary at
11,800\,km\,s$^{-1}$ at peak brightness, and found that the 
former are $\sim0.1$\,mag (on average) redder in $B-V$ than the 
latter.

Meanwhile, high-resolution spectral observations provide a powerful 
way to study absorption along the line of sight, both from the interstellar
medium and circumstellar material. Patat et al. (2007) found
a complex of \ion{Na}{1}~D lines that showed evolution in
SN~2006X (Wang et al. 2008). Two additional cases of time-variable
Na absorption are provided by
Blondin et al. (2009) and Simon et al. (2009).
Sternberg et al. (2014) found that in their sample with 
high-resolution spectra,
$\sim18$\% of SNe~Ia exhibit time-variable Na, indicating the presence of
circumstellar material and suggesting that it may be more common
than expected in SNe~Ia, though some objects do not show evolution
(e.g., SN~2014J; Graham et al. 2015).

Spectropolarimetry can be used to probe the geometry of
SNe~Ia (see Wang \& Wheeler 2008 for a review). The continuum polarization,
an indication of the photosphere's shape, was found to be quite
low in SNe~Ia, on the order of a few tenths of a percent
(H\"{o}flich 1991; Wang et al. 1997).
But for individual SNe~Ia, significant line polarization is
sometimes observed (e.g., Wang et al. 2003; Kasen et al. 2003).
Wang et al. (2007) also found a correlation between the degree of polarization
of \ion{Si}{2} $\lambda$6355 and the brightness decline rate.

Observationally, there are numerous efforts to discover SNe~Ia at very 
early times, which can benefit follow-up observations in many ways.
Recent examples of early-observed and well-studied SNe~Ia include SN~2009ig (Foley et al. 2012), 
SN~2011fe (Nugent et al. 2011; Li et al. 2011), SN~2012cg (Silverman et al. 2012a),
SN~2013dy (Zheng et al., 2013), iPTF13ebh (Hsiao et al. 2015), SN~2014J (Zheng et al. 2014;
Goobar et al. 2014; Graham et al. 2015), and ASASSN-14lp (Shappee et al. 2016);
like SN~2016coj discussed here, they were either discovered or detected shortly after exploding.

In 2011, the observing strategy for our Lick Observatory Supernova
Search (LOSS; Filippenko et al. 2001; Filippenko 2005; Leaman et al. 2011) 
with the 0.76\,m Katzman
Automatic Imaging Telescope (KAIT) was modified to monitor fewer
galaxies but at a more rapid cadence, with the objective of promptly
identifying very young SNe (hours to days after explosion).
In the past few years, this strategy has led to discoveries of all 
types of young SNe, where we define a SN to be ``young'' if there was a
KAIT nondetection 1--3 days before the first detection
or if it was spectroscopically confirmed to be within a few days 
after explosion.
SN~2012cg (Silverman et al. 2012a) was the
first case, followed by more than a dozen others (e.g., SN~2012ck, Kandrashoff
et al. 2012; SN~2012ea, Cenko et al. 2012; SN~2013ab, Blanchard et al. 2013;
SN~2013dy, Zheng et al. 2013; SN~2013ej, Dhungana et al. 2016;
SN~2013gh, Hayakawa et al. 2013; SN~2013fv, Kim et al. 2013a;
SN~2013gd, Casper et al. 2013; SN~2013gy, Kim et al. 2013b;
SN~2014C, Kim et al. 2014a; SN~2014J, though not discovered by KAIT,
but with KAIT early detections, see Zheng et al. 2014;
SN~2014ce, Kim et al. 2014b; SN~2014eh, Kumar et al. 2014; 
SN~2015N, Stegman et al. 2015a; SN~2015U, Shivvers et al. 2016;
SN~2015X, Hughes et al. 2015; SN~2015O, Ross et al. 2015;
SN~2015be, Stegman et al. 2015b; and SN~2016esw, Halevi et al.
2016\footnote{https://wis-tns.weizmann.ac.il//object/2016esw};).

SN~2016coj was another SN discovered by KAIT when very young,
merely $0.6 \pm 0.5$\,day after the fitted first-light time (FFLT).
Here we present the first 40 days of our optical photometric, low- and high-resolution spectroscopic, and spectropolarimetric follow-up observations 
and analysis of it.


\section{Discovery and Observations}\label{s:discovery}

SN~2016coj was discovered in an 18~s unfiltered KAIT image taken 
at 04:39:05 on 2016~May~28 (UT dates are used throughout this paper),
at $14.98\pm0.03$\,mag (close to the $R$ band; see Li et al. 2003).
It was reported to the Transient Name Server (TNS)
shortly after discovery by Yuk, Zheng, \& 
Filippenko\footnote{https://wis-tns.weizmann.ac.il//object/2016coj}
(see also Zheng et al. 2016).
We measure its J2000.0 coordinates to be 
$\alpha=12^{\mathrm{h}}08^{\mathrm{m}}06\farcs80$, 
$\delta=+65^{\circ}10\arcmin38\farcs2$, 
with an uncertainty of $0\farcs5$ in each coordinate.
SN~2016coj is $5\farcs0$ east and $10\farcs8$ north of
the nucleus of the host galaxy NGC~4125, which has
redshift $z=0.004523$
according to the NASA/IPAC Extragalactic Database
(NED\footnote{http://ned.ipac.caltech.edu/}), 
an early-type peculiar elliptical morphology 
(E6 pec; de Vaucouleurs et al. 1991),
and a stellar mass of $2.4 \times 10^{11}$\,\msun\ 
from its 3.6\,${\mu}m$ flux (Wilson et al. 2013).

\begin{figure}
\centering
\includegraphics[width=.49\textwidth]{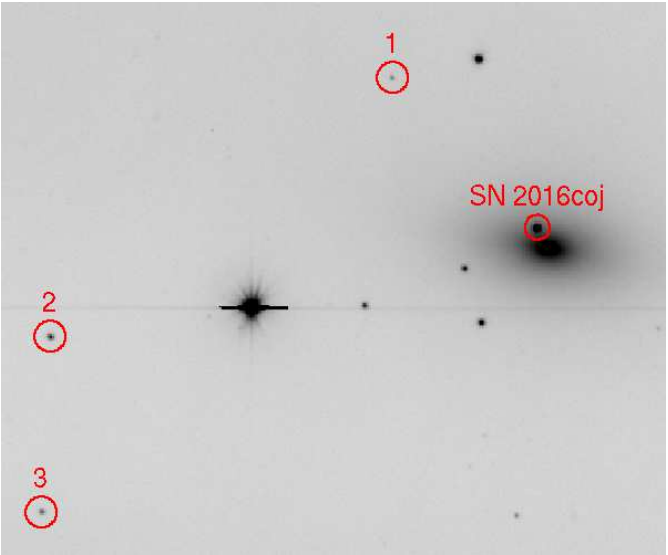}
\caption{KAIT unfiltered image showing the location of SN~2016coj. Three reference stars
  are also marked with circles.
  }
\label{Fig_kaitclear_findingchart}
\end{figure}

KAIT performed photometric follow-up observations of SN~2016coj with 
nearly daily cadence after discovery. The data were reduced using 
our image-reduction pipeline (Ganeshalingam et al. 2010).
We applied an image-subtraction procedure to remove host-galaxy light,
and point-spread-function photometry was then obtained using DAOPHOT 
(Stetson 1987) from the IDL Astronomy User's 
Library\footnote{http://idlastro.gsfc.nasa.gov/}.
The unfiltered instrumental magnitudes, which are found to be close 
to the $R$ band (Li et al. 2003),
are calibrated to local SDSS standards (see Figure \ref{Fig_kaitclear_findingchart}) transformed into Landolt $R$-band 
magnitudes\footnote{http://www.sdss.org/dr7/algorithms/sdssUBVRITransform.html\#Lupton2005}.
Here we publish our unfiltered photometry (Table 1).
We have also obtained a filtered data sequence, but we are still awaiting 
high-quality galaxy template images in those bands.

\begin{figure}
\centering
\includegraphics[width=.49\textwidth]{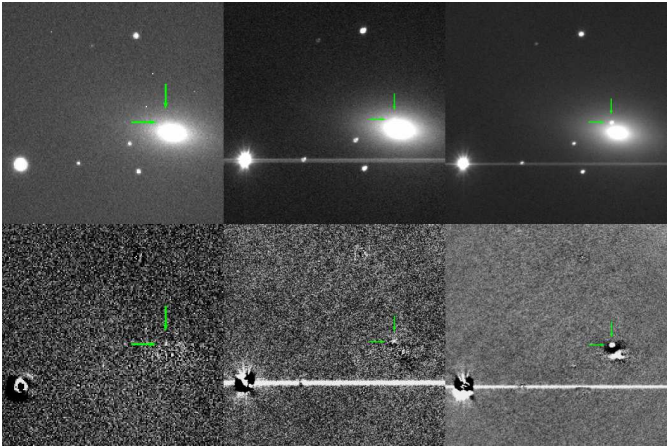}
\caption{Left panels: Arbour's image taken on May 23. Middle panels:
  KAIT unfiltered image taken on May 24. Right panels: KAIT unfiltered image
  taken on May 28. Upper panels show the original image and lower panels
  show the residual after subtraction; SN~2016coj is marked.
  }
\label{Fig_findingchart_subtracted}
\end{figure}

Interestingly, we find that SN~2016coj was detected in a 
KAIT prediscovery image taken at 04:30:35 on 
May 24 (see middle panels of Figure \ref{Fig_findingchart_subtracted})
with an unfiltered mag of $18.02 \pm 0.22$,
which means the SN had brightened $\sim3.0$ mag
in the following four days until it was discovered. 
In addition, an unfiltered prediscovery
detection was obtained at 21:48:57 on May 23, 6.7 hr earlier than
KAIT's first detection, by R. Arbour with a 0.35\,m $f/6$ 
Schmidt-Cassegrain reflector 
(see left panels of Figure \ref{Fig_findingchart_subtracted}).
Using a template image taken on 2016 April 5, we performed
the same subtraction and calibration methods as for the KAIT unfiltered images.
We find a SN unfiltered brightness of $18.06 \pm 0.42$\,mag, 
consistent with KAIT's first detection. 
A $\sim 5$\,mag detection before peak
magnitude, along with our analysis in the following section, 
confirms that SN~2016coj is one of the youngest SNe~Ia ever 
detected.

\begin{deluxetable}{lcccc}
 \tabcolsep 0.4mm
 \tablewidth{0pt}
 \tablecaption{Unfiltered Photometry of SN~2016coj
               \label{tab:prompt_spec_par}}
 \tablehead{ \colhead{MJD} & \colhead{UT} & \colhead{Mag} & \colhead{Error} & \colhead{From} }
 \startdata
\\              
57520.2370      &    May 12.2370    &   $>$19.6  & -    & KAIT          \\
57522.2513      &    May 14.2513    &   $>$19.4  & -    & KAIT          \\
57524.2240      &    May 16.2240    &   $>$19.4  & -    & KAIT          \\
57525.2479      &    May 17.2479    &   $>$19.3  & -    & KAIT          \\
57527.2708      &    May 18.2708    &   $>$19.2  & -    & KAIT          \\
57531.9092      &    May 23.9092    &      18.06 & 0.42 & R. Arbour     \\
57532.1877      &    May 24.1877    &      18.02 & 0.22 & KAIT          \\
57536.2694      &    May 28.2694    &      14.98 & 0.03 & KAIT          \\
57537.2196      &    May 29.2196    &      14.50 & 0.04 & KAIT          \\
57538.1827      &    May 30.1827    &      14.11 & 0.03 & KAIT          \\
57539.1814      &    May 31.1814    &      13.85 & 0.03 & KAIT          \\
57540.2518      &   June 01.2518    &      13.54 & 0.04 & KAIT          \\
57541.2022      &   June 02.2022    &      13.39 & 0.06 & KAIT          \\
57542.2075      &   June 03.2075    &      13.18 & 0.03 & KAIT          \\
57543.2120      &   June 04.2120    &      13.16 & 0.03 & KAIT          \\
57544.1995      &   June 05.1995    &      13.02 & 0.03 & KAIT          \\
57545.2245      &   June 06.2245    &      13.00 & 0.03 & KAIT          \\
57546.2200      &   June 07.2200    &      12.95 & 0.03 & KAIT          \\
57547.2051      &   June 08.2051    &      12.93 & 0.03 & KAIT          \\
57548.2197      &   June 09.2197    &      12.94 & 0.04 & KAIT          \\
57549.2162      &   June 10.2162    &      12.99 & 0.05 & KAIT          \\
57550.2638      &   June 11.2638    &      12.97 & 0.02 & KAIT          \\
57551.2201      &   June 12.2201    &      13.01 & 0.03 & KAIT          \\
57552.2516      &   June 13.2516    &      13.03 & 0.03 & KAIT          \\
57553.2115      &   June 14.2115    &      13.15 & 0.03 & KAIT          \\
57555.2189      &   June 16.2189    &      13.27 & 0.03 & KAIT          \\
57556.2284      &   June 17.2284    &      13.30 & 0.03 & KAIT          \\
57558.2110      &   June 19.2110    &      13.52 & 0.02 & KAIT          \\
57559.2229      &   June 20.2229    &      13.61 & 0.04 & KAIT          \\
57560.2189      &   June 21.2189    &      13.66 & 0.04 & KAIT          \\
57561.2097      &   June 22.2097    &      13.70 & 0.04 & KAIT          \\
\enddata
\end{deluxetable}

Two classification spectra of SN~2016coj were obtained shortly ($\sim3.7$\,hr)
after the SN was discovered ($\sim$ 5.0 days after the FFLT).
The spectra were taken with the Kast double spectrograph 
(Miller \& Stone 1993) on the Shane 3\,m telescope at 
Lick Observatory and the FLOYDS robotic spectrograph on the 
Las Cumbres Observatory Global Telescope Network (LCOGT; Brown et al. 2013)
2.0\,m Faulkes Telescope North on Haleakala, Hawaii.
We obtained nearly daily spectra of SN~2016coj
with different instruments including Kast, FLOYDS, the BFOSC spectrograph on the
2.16\,m telescope at Xinglong station of NAOC (China),
the Low Resolution Imaging Spectrometer (LRIS; Oke et al. 1995)
on the 10\,m Keck~I telescope, and the Kitt Peak 
Ohio State Multi-Object Spectrograph
(KOSMOS; Martini et al. 2014) on the KPNO Mayall 4\,m telescope.
Data were reduced following standard techniques for CCD processing and spectrum extraction using IRAF.
The spectra were flux calibrated through observations of appropriate spectrophotometric standard stars.
All Kast and LRIS spectra were taken at or near the parallactic angle
(Filippenko 1982) to minimize differential light losses caused by
atmospheric dispersion. 
Low-order polynomial fits to calibration-lamp spectra were used to
calibrate the wavelength scale, and small adjustments derived from 
night-sky lines in the target frames were applied. 
Flux calibration and telluric-band removal were done with our own
IDL routines; details are described by Silverman et al. (2012d).

We also obtained four epochs of Lick/Shane spectropolarimetry 
using the polarimetry mode of the Kast spectrograph on
May 30, June 8, June 16, and July 6. 
The spectra were observed at each of four waveplate angles
($0^\circ$, $45^\circ$, $22.5^\circ$, and $67.5^\circ$) with 
several waveplate sequences coadded to improve the signal-to-noise ratio 
(S/N). Each night, both low- and high-polarization standard stars 
were also observed in order to calibrate the data.
All of the spectropolarimetric reductions and calculations 
follow the method described by Mauerhan et al. (2015),
and the polarimetric parameters are defined in the same manner 
(Stokes parameters $q$ and $u$, debiased polarization $P$, and 
sky position angle $\theta$). 

In addition, we observed SN\,2016coj on May 31, June 2, 4, and 6 with
the 2.4\,m Automated Planet Finder (APF) telescope at Lick Observatory.
The APF hosts the Levy Spectrograph, a high-resolution optical 
echelle spectrograph
with resolution $R$(5500\,\AA) $\approx$ 110,000 with a slit width of 
1\arcsec\ (Vogt et al. 2014).
At each epoch we obtained three 1800\,s spectra with the M decker
(1\farcs0 wide, 8\farcs0 long to allow for background subtraction),
reduced the data with a custom pipeline, and corrected for the redshift
of the host galaxy ($z=0.004523$) and for the barycentric velocity
($\sim -15$\,km\,s$^{-1}$). Because the apparent magnitude 
(peaking at $\sim13$ mag) of SN\,2016coj is a bit faint for APF,
the S/N of our spectra was $\lesssim10$ at best, 
significantly lower than obtained with APF spectra of the bright
(peak $\sim10$\,mag), nearby SN\,Ia 2014J (Graham et al. 2015).

\begin{figure}
\centering
\includegraphics[width=.49\textwidth]{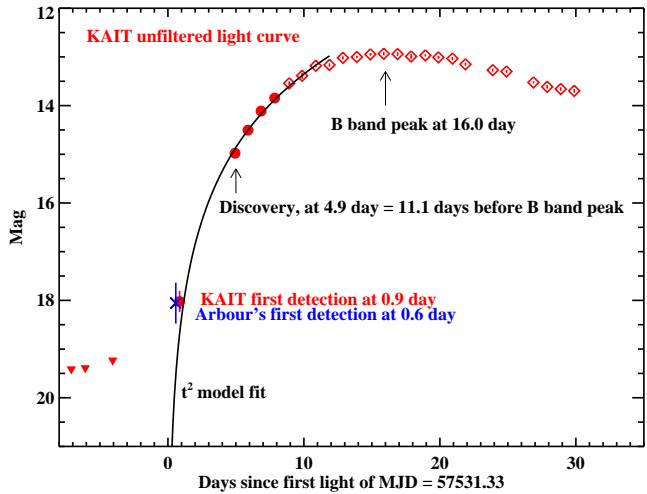}
\caption{KAIT unfiltered (red) light curve of SN~2016coj.
  The solid black line is the $t^{2.0}$ model fit to the red solid circles.
  The blue cross marks the earliest detection at $0.6\pm0.5$\,d after the fitted first light observed
  by R. Arbour. Red triangles show the KAIT upper limits before explosion.
  }
\label{Fig_lcfit}
\end{figure}


\section{Analysis and Results}\label{s:analysis}

\subsection{Light Curve}\label{ss:lightcurves}

Figure \ref{Fig_lcfit} shows our unfiltered light curve of SN~2016coj.
In order to determine the first-light time $t_0$ (note that the SN may
exhibit a ``dark phase"),
one can assume that the SN luminosity scales as the
surface area of the expanding fireball, and therefore increases
quadratically with time ($L \propto t^{2}$, commonly known as the
$t^2$ model; Arnett 1982; Riess et al. 1999). The $t^2$ model fits well for several
SNe~Ia with early-time observations (e.g., SN~2011fe, Nugent et al. 2011;
SN~2012ht, Yamanaka et al. 2014). Some studies also adopt a $t^n$ model
($n$ varies from $\sim1.5$ to $\sim3.0$; e.g., Conley et al. 2006;
Ganeshalingam et al. 2011; Firth et al. 2015). Interestingly, Zheng et al. (2013, 2014)
use a broken-power-law model to estimate the first-light time of SN~2013dy and SN~2014J.
However, since our early-time photometric coverage of SN~2016coj is not as good as that of SN~2013dy and SN~2014J, 
we simply apply the $t^2$ model to fit the KAIT
unfiltered data for the first few days (red solid circles in Figure \ref{Fig_lcfit}); thereafter, the light curve starts deviating from the $t^2$ model.
We also exclude Arbour's unfiltered detection (blue cross), considering
the different response curve compared to the KAIT unfiltered data:
Arbour's unfiltered band is closer to $V$ (see Botticella et al. 2009), while
KAIT's is closer to $R$ (see Li et al. 2003).

We find that the best $t^2$ model fit gives the first-light time to be
MJD $=57531.33\pm0.50$, around May 23.33. Here the uncertainty (not including the ``dark phase") is
estimated by calculating the reduced $\chi^2$ ratio with the minimum reduced $\chi^2$
at 90\% confidence level, when $t_0$ changes around the best-fitted value while all the
other parameters are fixed with the best-fitted value.
Note that the uncertainty does not include any systematic error caused by the
$t^2$ model fitting. For example, if we include (or exclude) one data point
before and after the dataset we used, the best-fit first-light time
deviates $-0.4$ to 1.0 days from the above first-light time.
Therefore, there could be a systematic error of up to 1.0\,d from this method,
which we did not include in the following analysis.
Our results show that the first detection (from an image by R. Arbour) was merely $0.6\pm0.5$\,d after first light,
or 0.9\,d from KAIT's first detection on May 24.
This makes SN~2016coj one of the earliest detected SNe~Ia ---
slightly later than SN~2013dy ($\sim2.4$\,hr after first light; Zheng et al. 2013)
and SN~2011fe ($\sim11.0$\,hr after first light; Nugent et al. 2011), but similar to
SN 2009ig ($\sim17$\,hr after first light; Foley et al. 2012).

Applying a low-order polynomial fit, we find that SN~2016coj reached a 
peak magnitude of $12.91\pm0.03$ at MJD $=57547.31$ in KAIT unfiltered data.
Although we do not present $B$-band data because no $B$-band template image
is currently available,
the fit allows us to determine the $B$-band peak time: MJD $=57547.35$, 
similar to the result with unfiltered data.
This means SN~2016coj was discovered only 4.9\,d after the fitted first 
light, or 11.1\,d before maximum light.

The distance modulus of the host galaxy NGC~4125 is quite
uncertain owing to different measurements given in NED.
However, some of them are outdated, or adopted an inappropriate H$_0$ value.
The one with the smallest uncertainty (and also the latest estimate)
is $31.90\pm0.14$\,mag (Tully et al. 2013),
which was based on H$_0=74.4$\,km\,s$^{-1}$\,Mpc$^{-1}$, quite close 
to the current widely accepted value of H$_0 \approx 70$. 
We therefore adopt this distance for the following ananlysis.
With $E(B-V)_\textrm{MW} = 0.02$\,mag (Schlafly et al. 2011) and
very small (even negligible) host-galaxy extinction 
(see \S\ref{ss:spectra} and \S\ref{ss:highresspectra}), 
this implies SN~2016coj has $M_R = -19.0\pm0.2$\,mag at peak brightness.
Our preliminary measurement of $B$-band data, assuming
the host background contamination is small,
shows that the $B$-band peak is $\sim13.1\pm0.1$\,mag
and $\Delta m_{15}(B)=1.25\pm0.12$\,mag.
This gives $M_B\approx -18.9\pm0.2$\,mag, but we
expect $M_B\approx -19.1$\,mag from the Phillips (1993) 
relation with the above value of $\Delta m_{15}(B)$;
thus, SN~2016coj is a normal-brightness SN~Ia.
Its $\Delta m_{15}(B)=1.25\pm0.12$\,mag
is also typical of normal SNe~Ia
(see also \S\ref{ss:classification} for the spectral classification). 

\subsection{Optical Spectra}\label{ss:spectra}

\begin{figure*}
\centering
\includegraphics[width=.98\textwidth]{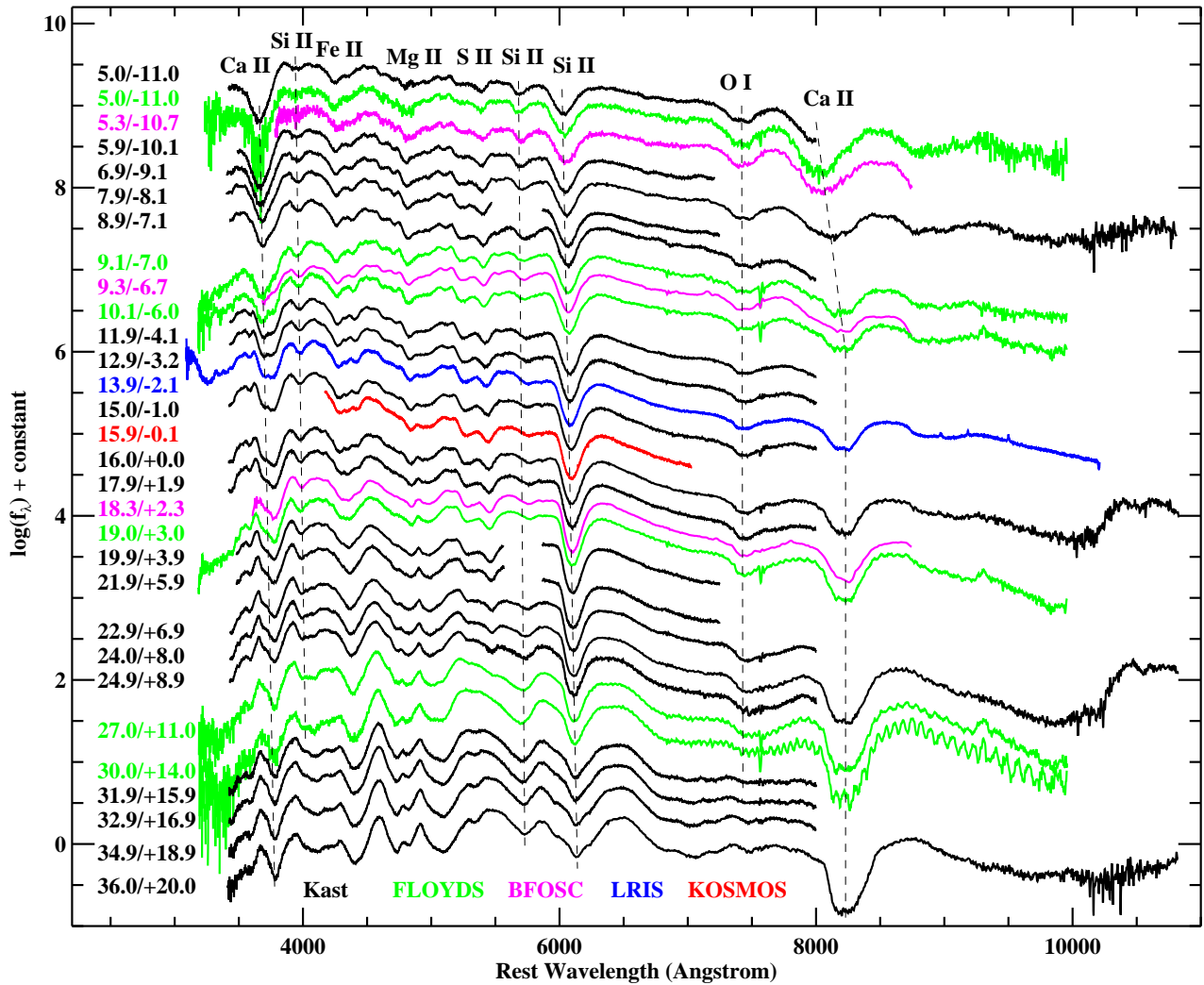}
\caption{Spectral sequence of SN~2016coj over the first month after discovery.
  Each spectrum is labeled with its age relative to both the 
  fitted first light and to the $B$-band maximum light.
  Some major spectral features are labeled at the top.
  Spectra taken by different instruments are shown in different colors.
  Three Lick/Kast spectra have no coverage around 5500--5900\,\AA; those
  sections are left blank. Dashed lines are meant to help guide the eye
  when examining absorption features.
  }
\label{Fig_spectra}
\end{figure*}

We obtained optical spectra of SN~2016coj nearly daily for a month
(Fig.~\ref{Fig_spectra}),
sometimes obtaining multiple spectra in a given night.

We first examine the \ion{Na}{1}~D absorption feature, which is often converted into reddening but with
large scatter over the empirical relationship (Poznanski et al. 2011),
in several of our high-S/N spectra.
The absorption is not clearly detected at both the \ion{Na}{1}~D 
rest-frame wavelength and the redshifted wavelength of SN~2016coj.
However, there appears to be a weak absorption feature consistent with 
the rest-frame
\ion{Na}{1}~D wavelength. If real, this could be caused by a Milky Way component,
which has $E(B-V)_\textrm{MW} = 0.02$\,mag according to Schlafly et al. (2011).
Since we do not detect similar absorption at the redshifted wavelength of SN~2016coj,
we can put an upper limit of $E(B-V)\lesssim0.02$\,mag of host-galaxy extinction.
However, if the weak absorption feature is caused by noise instead of Milky Way gas,
we can determine an upper limit on $E(B-V)$ through comparison with our
spectra of SN~2013dy (Zheng et al. 2013), where we clearly
detect the \ion{Na}{1}~D absorption with the same instrument setting. For SN~2013dy,
the equivalent width ($W_\lambda$) is $\sim0.5$\,\AA\
from both the Milky Way and host galaxy, giving $E(B-V)=0.15$\,mag. 
Our similar-quality data on SN~2016coj should allow a detection of 1/3 (or less) 
of \ion{Na}{1}~D absorption if it exists,
yielding an upper limit of $E(B-V)\lesssim0.05$\,mag of host-galaxy extinction.
Lastly, we also estimate a
$3\sigma$ upper limit on the $W_\lambda$ of an undetected feature in
a spectrum using the equation presented by
Leonard (2007):
$W_\lambda(3\sigma) = 3\Delta\lambda\ \Delta I \sqrt{W_{\rm line} / \Delta\lambda} \sqrt{1 / B},$
\noindent where $\Delta \lambda$ is the spectral resolution element (in \AA),
$\Delta I$ is the 1$\sigma$ root-mean-square fluctuation of the flux around a
normalized continuum level, $W_{\rm line}$ is the full-width at half-maximum
intensity (FWHM) of the expected line feature, and $B$ is the number of bins 
per resolution element.
For our high-S/N Kast spectra, we measure
$\Delta\lambda\approx 4.0$\,\AA, $\Delta I\approx 0.015$, $W_{\rm line}\approx
12.0$\,\AA, and $B=1$, which gives $W_\lambda(3\sigma)\approx 0.3$\,\AA,
and $E(B-V)\lesssim 0.09$\,mag of host-galaxy extinction.

All of the above suggests that the host-galaxy extinction of SN~2016coj is likely to
be very small, consistent with the nondetection of \ion{Na}{1}~D absorption
in our high-resolution spectra (see \S\ref{ss:highresspectra}).
However, note that since the \ion{Na}{1}~D vs. extinction relation has large scatter,
even a nondetection of \ion{Na}{1}~D does not fully exclude
the possibility that there may be some dust along the
SN line of sight.

The spectra show absorption features from ions typically seen in SNe~Ia including
\ion{Ca}{2}, \ion{Si}{2}, \ion{Fe}{2}, \ion{Mg}{2}, \ion{S}{2}, and \ion{O}{1}.
We do not find a clear \ion{C}{2} feature (e.g., Zheng et al. 2013), which is 
found in over one-fourth of all SNe~Ia (e.g., Parrent et al. 2011; Thomas et al. 2011;
Folatelli et al. 2012; Silverman et al. 2012b).
Strong absorption features of \ion{Si}{2}, including \ion{Si}{2} $\lambda$4000,
\ion{Si}{2} $\lambda$5972, and \ion{Si}{2} $\lambda$6355, are clearly seen in all spectra.
The \ion{Si}{2} $\lambda$5972 feature in SN~2016coj is quite strong relative to those
in SN~2012cg and SN~2013dy, though it is relatively small if compared with a
large SN~Ia sample (see Silverman et al. 2012c).

We measure the individual line velocities from the minimum of the absorption features
(see Silverman et al. 2012c, for details) and show them in Figure \ref{Fig_spec_velocity}.
The velocities all \ion{Si}{2} features decrease from $\sim13,000$--15,000\,\kms\ 
at discovery to $\sim11,000$--13,000\,\kms\ around maximum light, and they continue to
decrease thereafter. 

\begin{figure}
\centering
\includegraphics[width=.48\textwidth]{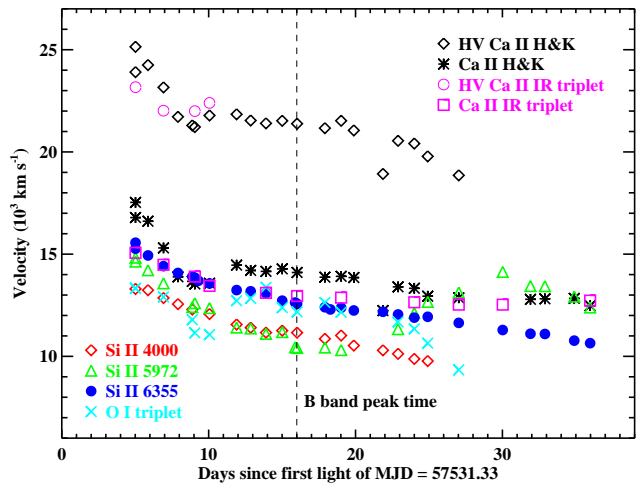}
\caption{The expansion velocity evolution of different lines measured
  from the spectra of SN~2016coj. The black dashed line marks the 
  time of $B$-band maximum light.
  }
\label{Fig_spec_velocity}
\end{figure}

In addition to the usual photospheric-velocity feature (PVF) of \ion{Ca}{2} H\&K,
SN~2016coj exhibits a high-velocity feature (HVF; e.g., Mazzali et al. 2005;
Maguire et al. 2012; Folatelli et al. 2013; Childress et al. 2014;
Maguire et al. 2014; Silverman et al. 2015) in nearly all of the early-time
spectra.
This HVF appears to be detached from the rest of the photosphere,
with a velocity of $\sim25,000$\,\kms\ at discovery and slowing down to
$\sim20,000$\,\kms\ at $\sim8$\,d after the fitted first-light time.
The HVF feature of \ion{Ca}{2} H\&K stays for a long time, being distinct
until roughly age +11\,d; thereafter, it is a high-velocity shoulder
of the \ion{Ca}{2} H\&K absorption.

A \ion{Ca}{2} near-infrared (NIR) triplet HVF is also found in the first few
spectra that covered the wavelength range before maximum light, and the velocity of
$\gtrsim22,000$\,\kms\ is in good agreement with that of the \ion{Ca}{2} H\&K HVF 
at early times, though it is slightly smaller in the first-epoch spectrum.
Such HVFs are seen in a few other well-observed SNe~Ia, including SN~2005cf (Wang et al. 2009) and
SN~2012fr (e.g., Maund et al. 2013; Childress et al. 2013; Zhang et al. 2014).
However, in SN~2016coj, the \ion{Ca}{2} NIR triplet HVF becomes weaker around peak 
brightness, and it completely disappears $\sim8$\,d later and thereafter. 
This is different
from the \ion{Ca}{2} H\&K HVF, which is seen for a much longer time.
It is not obvious why the HVF of the \ion{Ca}{2} NIR triplet goes away after
peak brightness while the HVF of \ion{Ca}{2} H\&K persists. 
One possibility is that the apparent HVF
of \ion{Ca}{2} H\&K after peak could actually be \ion{Si}{2} $\lambda$3858
(e.g., Foley 2013).
In fact, it is possible that the early-time apparent HVF of \ion{Ca}{2} H\&K could be
a mixture of \ion{Si}{2} $\lambda$3858 (including both the HVF and PVF) 
plus the true HVF of \ion{Ca}{2} H\&K. If so, the velocity of the 
\ion{Ca}{2} H\&K HVF could be smaller than that shown in 
Figure \ref{Fig_spec_velocity}, and thus more consistent with
the velocity of the \ion{Ca}{2} NIR triplet HVF, but this case is too 
complicated to verify.

One note about the \ion{O}{1} triplet feature is that we adopted only one component
in our fit. However, our early-time spectra before peak brightness reveal that the
\ion{O}{1} triplet has a double absorption profile. Following the Zhao et al. (2016)
method to fit the \ion{O}{1} triplet with both HVF and PVH (Zhao et al. also adopted
a second, faster HVF, but that is not clear in SN~2016coj), we find an HVF 
\ion{O}{1} triplet velocity of $\sim16,000$\,\kms\ and a PVF \ion{O}{1} triplet
velocity of $\sim12,000$\,\kms. The HVF velocity is smaller than that of both
\ion{Ca}{2} H\&K and the \ion{Ca}{2} NIR triplet. If the HVF really exists in
the \ion{O}{1} triplet, it suggests that the oxygen in the outer layers is not 
completely burned (see Zhao et al. 2016).

The strong absorption of Si~II $\lambda$6355 is commonly used to
estimate the photospheric velocity. As shown in Figure \ref{Fig_spec_velocity}, 
the Si~II $\lambda$6355 velocity of SN~2016coj decreases rapidly from 
$\sim15,500$\,\kms\ at discovery to $\sim12,600$\,\kms\ around peak brightness,
and then slowly decreases to $\sim11,600$\,\kms\ at +11.0\,d after peak. 
A velocity of $\sim12,600$\,\kms\ at peak brightness is $\sim1500$\,\kms\
higher than average in SNe~Ia (e.g., Wang et al. 2013; $>2.5\sigma$ away from
the mean of their SN~Ia velocity distribution fitted with a Gaussian).
Here, we compare the photospheric velocity measurement of SN~2016coj with 
the three well-observed SNe~Ia
2009ig (Foley et al. 2012; Marion et al. 2013), 2012cg (Silverman et al. 2012a), and
2013dy (Zheng et al. 2013). Note that while both SN~2012cg and SN~2009ig have an HVF
identified for \ion{Si}{2} $\lambda$6355, we consider only the photospheric component.

\begin{figure*}
\centering
\includegraphics[width=.24\textwidth]{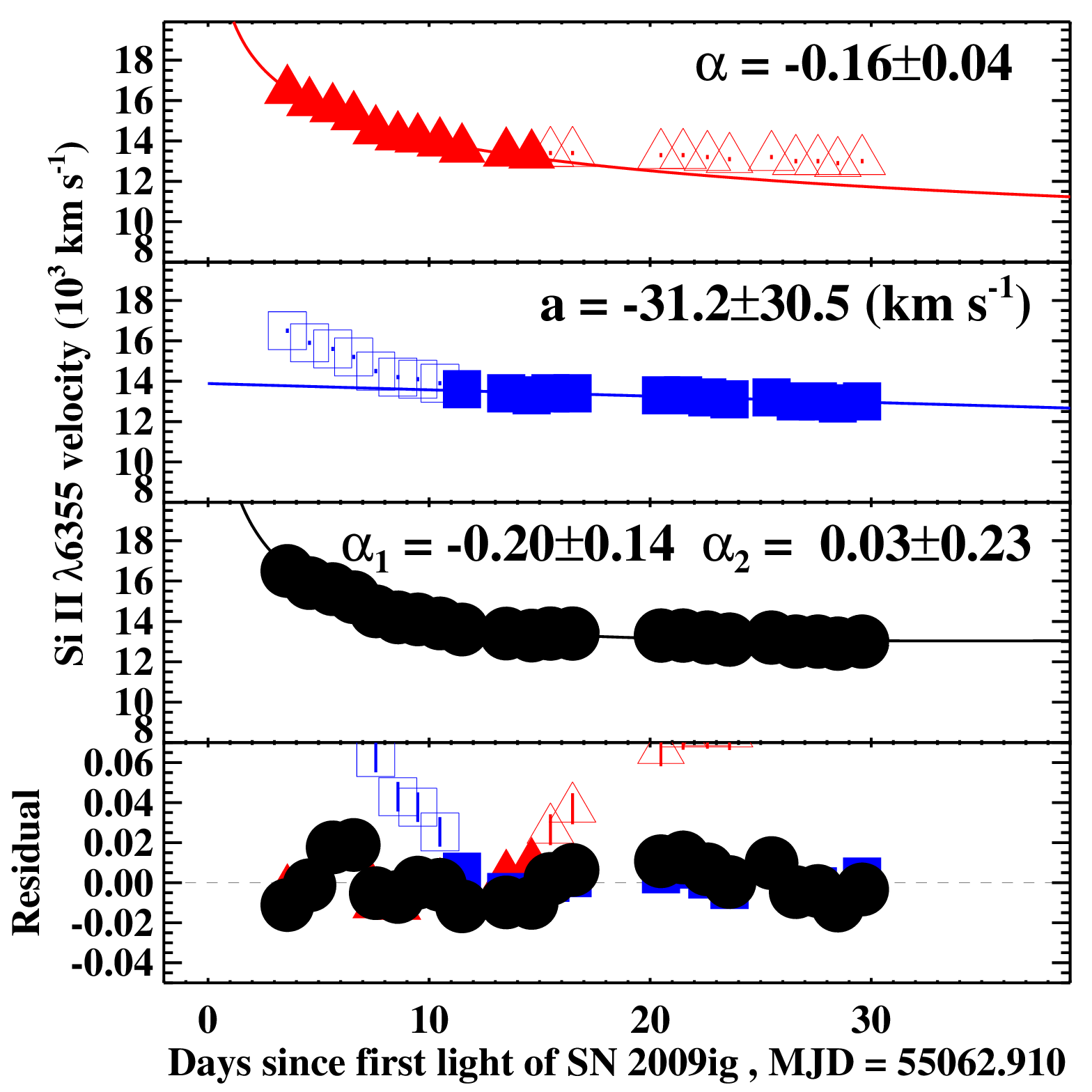}
\includegraphics[width=.24\textwidth]{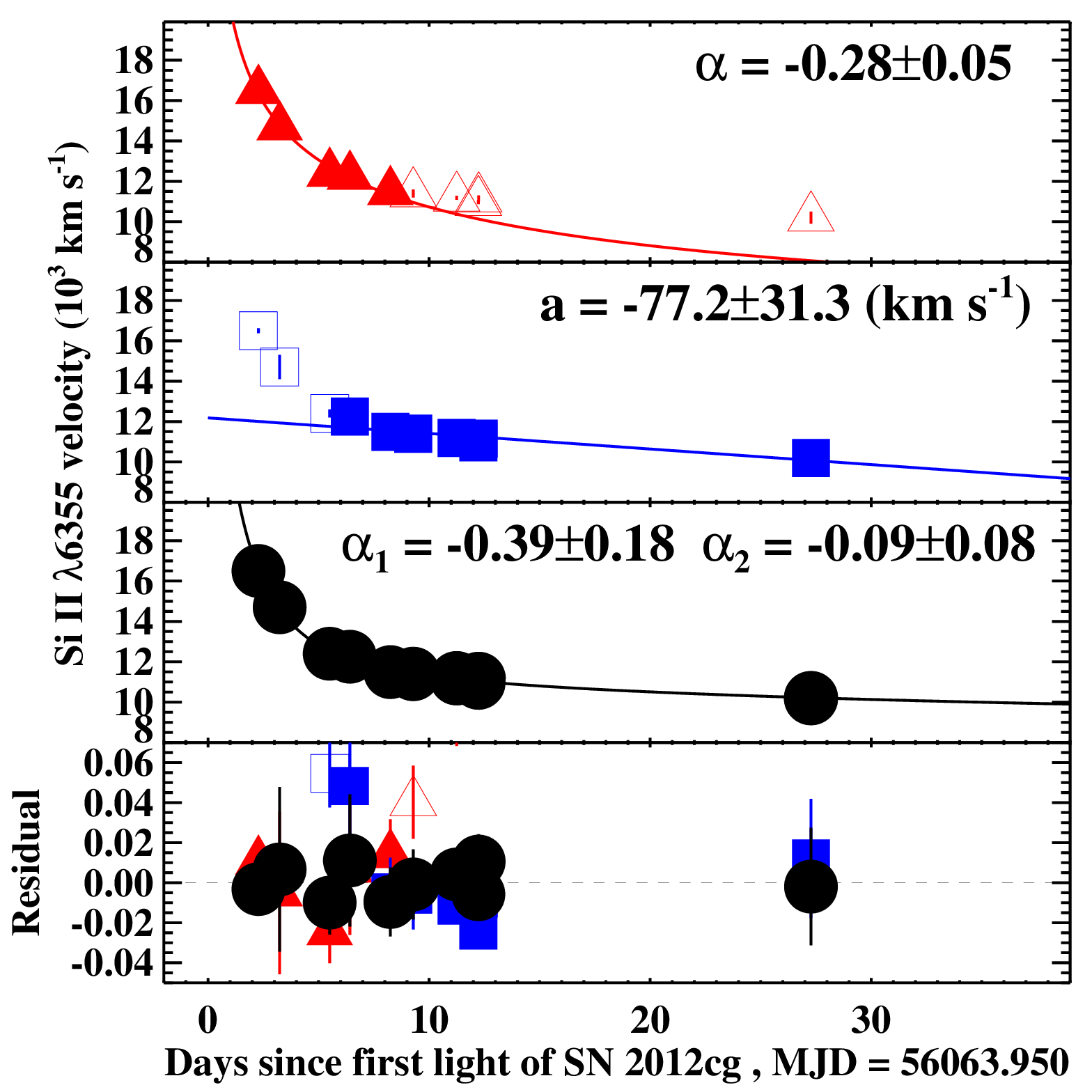}
\includegraphics[width=.24\textwidth]{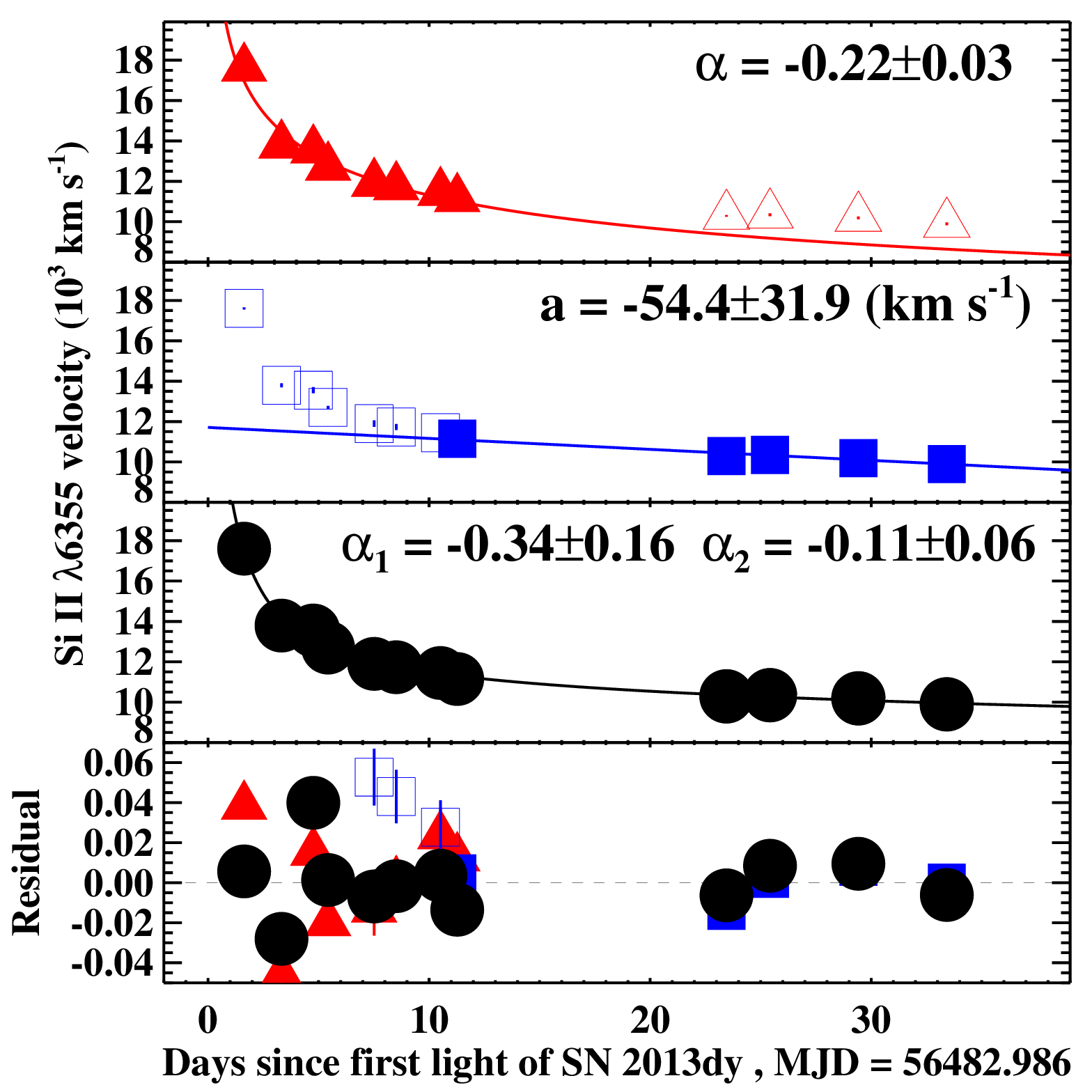}
\includegraphics[width=.24\textwidth]{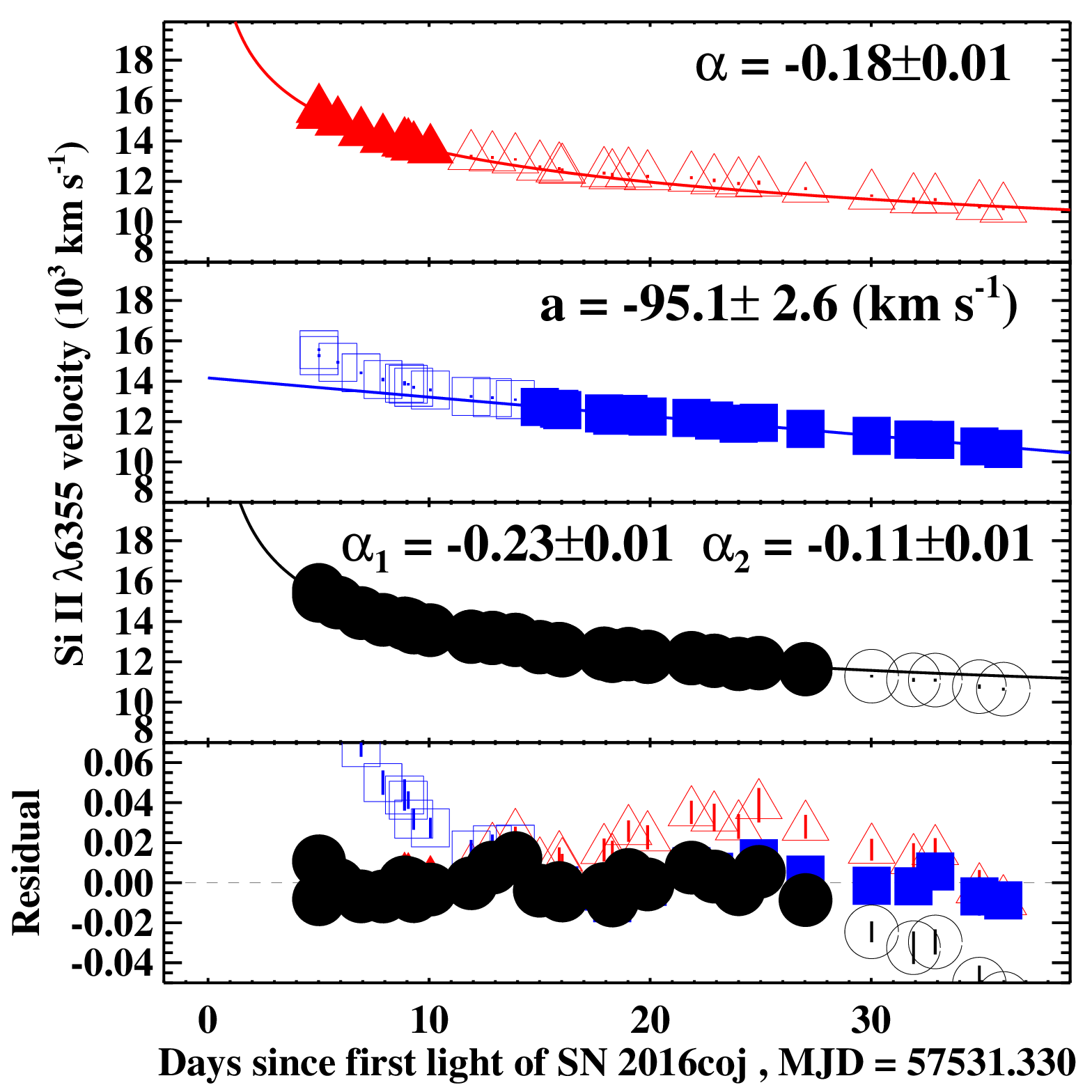}
\caption{Photospheric velocity (measured from the strong Si~II $\lambda$6355 absorption)
         evolution of SN~2016coj (right panel), compared to those of the well-observed
         SN~2009ig (left panel), SN~2012cg (middle-left panel), and SN~2013dy
         (middle-right panel). For all SNe~Ia, the top panels show the result of 
         a power-law function fit to the early-time data, the middle-top panels 
         display the result of a linear function fit to the later-time data,
         the middle-bottom panels give the result of a broken-power-law function 
         fit to all the data, and the bottom panels show the residuals for each 
         fit. Solid points are included in the fitting while open points are not.}
\label{Fig_spec_velocity}
\end{figure*}

Figure \ref{Fig_spec_velocity} displays the photospheric velocity evolution over
time for the four SNe~Ia.
Overall, the photospheric velocity evolution is similar to the evolution 
seen in most SNe~Ia (e.g., Benetti et al. 2005; Foley et al. 2011b; Silverman et al. 2012c):
the velocity drops rapidly at early times (within the first week after explosion),
and then slowly but steadily decreases thereafter. For each of these four SNe, we 
consequently try to fit the early-time velocities (typically within 10\,d after first 
light) with a power-law function, $v = C_1 t'^{\alpha}$,
where $t'$ is the time after first light ($t_0$); the results 
are shown in the top 
panel of Fig.~\ref{Fig_spec_velocity} for each SN. This is very similar to the method
Silverman et al. (2015, Fig.~12) adopted, but they used a natural exponential function
to fit the velocities before +5\,d after peak brightness and also obtained reasonable 
fitting results. In fact, Piro \& Nakar (2013, Eq.~13) mathematically show that the 
photospheric velocity could decay as a power law at early times.
For the later velocities (typically $>10$\,d after first light),
we then fit them with a linear function,
$v = a t' + C_2$ (results shown in the middle-top panel); Silverman et al. (2012c, Fig.~5)
also use the same method to fit their data around peak brightness.
As seen in Figure \ref{Fig_spec_velocity}, both the
power-law function and the linear function can fit the corresponding 
data well, but only in their respective regimes --- early-time data for the 
power-law function and later-time data for the linear function.


As with the early-time light-curve (Zheng et al. 2013, 2014),
we find that a broken-power-law function is useful 
for fitting the photospheric velocity evolution; a low-index
power-law function approximates the linear function found at
late times. Specifically,
\begin{equation}
v = A \left(\frac{t'}{t_b}\right)^{{\alpha}_1} \Big{[} 1 +
\left(\frac{t'}{t_b}\right)^{s({\alpha}_1-{\alpha}_2)}\Big{]}^{-1/s},
\label{Formula_bkn_vc}
\end{equation}
where $v$ is the photospheric velocity, $A$ is a scaling constant,  $t'$ is the time after first light ($t_0$),
$t_b$ is the break time, ${\alpha}_1$ and ${\alpha}_2$ are the two power-law indices before
and after the break (respectively), and $s$ is a smoothing parameter.
We apply this broken-power-law function to the entire dataset 
of photospheric velocities for all four SNe until about a month 
after the explosion.
Our fitting results (we fixed $s$ to be $-10$) are listed in Table 2 and
shown in the middle-bottom panels in Figure \ref{Fig_spec_velocity}.

\begin{deluxetable*}{lc|c|ccc}
 \tabcolsep 0.4mm
 \tablewidth{0pt}
 \tablecaption{Photospheric Velocity Fitting Results}
  \tablehead{\colhead{} & \colhead{ $\alpha$} & \colhead{a} & \colhead{$\alpha_1$} & \colhead{ $\alpha_2$} & \colhead{reduced $\chi^2$} }
\startdata
SN &  power law & linear$^a$ & \multicolumn{3}{c} {broken power law$^b$}      \\
\hline
SN~2009ig & -0.16$\pm$0.04 & -32$\pm$31 & -0.20$\pm$0.14 &  0.03$\pm$0.23  &  0.04 \\
SN~2012cg & -0.28$\pm$0.05 & -77$\pm$32 & -0.39$\pm$0.18 & -0.09$\pm$0.08  &  0.04 \\
SN~2013dy & -0.22$\pm$0.03 & -54$\pm$32 & -0.34$\pm$0.16 & -0.11$\pm$0.06  &  0.14 \\
SN~2016coj& -0.18$\pm$0.01 & -95$\pm$3  & -0.23$\pm$0.01 & -0.11$\pm$0.01  &  1.72 \\
\enddata
\tablenotetext{a}{In units of \kms\,d$^{-1}$.}
\tablenotetext{b}{The smoothing parameter $s$ was fixed to $-10$ during fitting, and the very small reduced $\chi^2$ for some SNe is largely caused by overestimating the velocity uncertainty.}
\end{deluxetable*}

The power-law indices from both the power-law fitting ($\alpha$) and broken-powerlaw fitting ($\alpha_1$) at early times are
consistent with the value of $-0.22$ adopted by Piro \& Nakar (2014) when fitting three SNe~Ia (SNe~2009ig, 
2011fe, and 2012cg), and are also the value adopted by Shappee et al. (2016) when fitting ASASSN-14lp.
The index from the broken power law ($\alpha_1$) is slightly steeper than that from the power law ($\alpha$).
At late times (around maximum light) with linear fitting, the rate of velocity decrease from the fitting is
slightly larger than the average rate of $-38$\,km\,s$^{-1}$\,d$^{-1}$ 
found by Silverman et al. (2012c) for a large sample of SNe~Ia.

Overall, the broken-power-law function can fit the photospheric velocity evolution well for all four SNe
until a month after explosion (see the small residuals at the bottom panel of
Figure \ref{Fig_spec_velocity}
and the reduced $\chi^2$ given in Table 2).
This function also has the potential to fit the photospheric velocity evolution
of most other SNe~Ia as well, given that most SNe~Ia have very similar velocity evolution
(e.g., Silverman et al. 2012b, 2012c). High-cadence spectroscopy
is required to verify this, especially at early times.
However, currently it remains unclear whether there is a good physical explanation behind the fitting;
Piro \& Nakar (2013) show that the photospheric velocity
could decay as a power law at early times, but our broken-power-law function fitting
extends to a much later time. 

\subsection{Classification}\label{ss:classification}


We use the SuperNova IDentification code (SNID; Blondin \& Tonry 2007) to spectroscopically classify SN~2016coj.
For nearly all of the spectra presented here, we find that SN~2016coj is spectroscopically similar to many
normal SNe~Ia.
Compared to SN~1992A ($M_B = -18.79$\,mag and $\Delta m_{15}(B) = 1.47$\,mag; Della Valle et al. 1998)
and SN~2002er ($M_B = -19.35$\,mag and $\Delta m_{15}(B) = 1.33$\,mag; Pignata et al. 2004), for example,
SN~2016coj has similar spectra, absolute magnitude, and $\Delta m_{15}(B)$.
Another spectroscopic comparison is the so-called \ion{Si}{2} ratio, $\Re$(\ion{Si}{2}) (the ratio of \ion{Si}{2} $\lambda$5972 to \ion{Si}{2} $\lambda$6355),
defined by Nugent et al. (1995) using the {\it depths} of spectral features and later by Hachinger et al. (2006)
using their pseudo-equivalent widths. Hachinger et al. (2006) found a good correlation
between the $\Re$(\ion{Si}{2})
and $\Delta m_{15}(B)$ (see their Figure 13). We measure SN~2016coj to have
$\Re$(\ion{Si}{2}) $= 0.11\pm0.4$, with $\Delta m_{15}(B) = 1.25$\,mag, placing SN~2016coj
in the normal SN~Ia region in Figure 13 of Hachinger et al. (2006), very close to SN~2002er.
Thus, we conclude that SN~2016coj is a spectroscopically normal SN~Ia, consistent with the
photometric analysis given in \S\ref{ss:lightcurves}.

\subsection{Spectropolarimetry}\label{ss:spectropolarimetry}

\begin{figure}
\centering
\includegraphics[width=.45\textwidth]{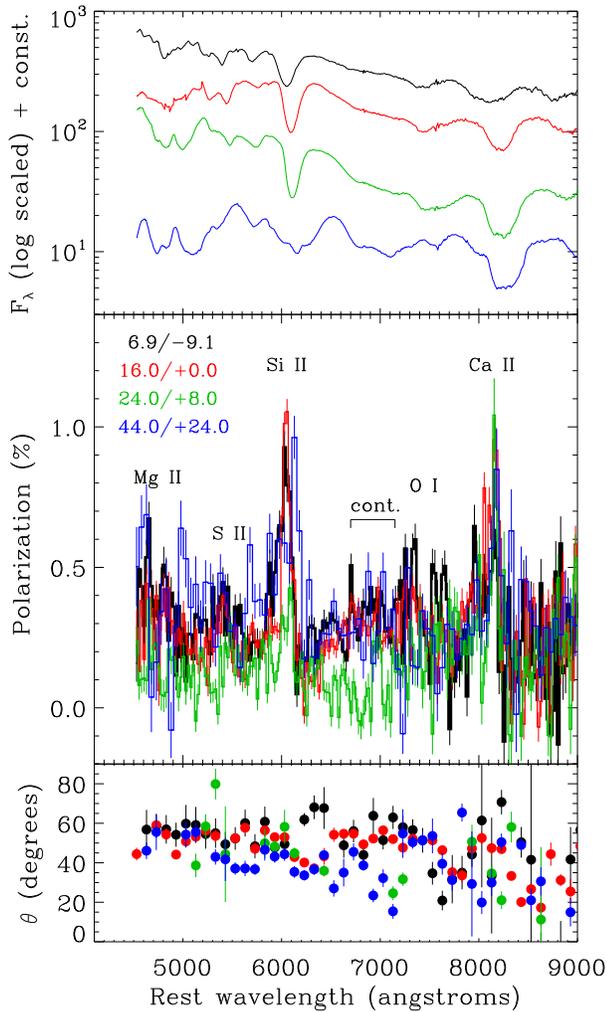}
\caption{Four epochs of spectropolarimetry of SN~2016coj. Top panel: observed total-flux spectrum, color coded for each epoch.
         Middle panel: Debiased polarization ($P$), with several major features labeled.
	 Bottom panel: Position angle ($\theta$) for the corresponding epochs. $\theta$ is underdetermined
	 where $P/\sigma P>1.5$; those points are omitted.
	 }
\label{Fig_specpol}
\end{figure}
\subsubsection{Interstellar and Instrumental Polarization}
The interstellar polarization (ISP) appears to be low in the direction of SN~2016coj. Indeed, the estimated value of $E(B-V)=0.02$\,mag indicates that the extinction from the Milky Way and host galaxy are not substantial; a small contribution from ISP is thus to be expected. According to Serkowski et al. (1975), an upper limit to the ISP is given by $9\times E(B-V)$, which implies $P_{\rm ISP}<0.18\%$ for SN~2016coj. To obtain a direct estimate of the Galactic component of ISP, we observed three Galactic stars in the vicinity of the SN position: HD\,104436 (A3\,V), HD\,106998 (A5\,V), and HD\,108907 (M3\,III). We measure respective $V$-band polarization and $\theta$ values of $P=(0.12\%, 0.09\%, 0.09\%)$ and $\theta=(36^{\circ}, 34^{\circ}, 30^{\circ})$. Under the reasonable assumption of low intrinsic polarization for these stars, the resulting average values of $P\approx0.1$\%, $\theta \approx33^{\circ}$ confirm the low Galactic polarization. Furthermore, the lack of \ion{Na}{1}~D absorption lines in our low- and high-resolution spectra (see \S\ref{ss:spectra} and \S\ref{ss:highresspectra}) indicates low extinction from the early-type host galaxy, and thus implies that the host ISP is probably even lower than the small Galactic value.

The instrumental polarization of the Kast instrument is also low. Measurements of the low-polarization standard star BD$+$33\,2642 at each epoch indicate an average $V$-band polarization of $\sim0.15$\%, with a standard deviation of 0.05\% between all four epochs; the average value is consistent with that reported by Schmidt et al. (1992) for this star, which indicates that the low level is intrinsic to the source and that Kast contributes an insignificant amount of instrumental polarization to the measurements. The standard deviation is near the systematic uncertainty level we typically experience using the spectropolarimetry mode of Kast. Our observations therefore constrain the average instrumental polarization to $<0.05\%$. Based on the low values of ISP and instrumental polarization, we move forward without attempting to subtract their minor contributions from the data.

\subsubsection{Intrinsic Polarization}

Our spectropolarimetry results are shown in Figure~\ref{Fig_specpol} and the integrated broadband measurements are listed in Table\,3. On day 6.9, the source exhibits weak polarization in the continuum at a level of $\sim0.3$\%, integrated over the wavelength range 6700--7150\,{\AA}. This is consistent with the weak levels of continuum polarization that are typically associated with SNe~Ia (Wang \& Wheeler 2008), although we note that some fraction of the polarization, perhaps half, could potentially be contributed by ISP. Strong polarization is exhibited across prominent line features, particularly \ion{Si}{2} $\lambda$6355 and the \ion{Ca}{2} NIR triplet, at levels of $\sim0.9$\% and $\sim0.6$\%, respectively. The \ion{Ca}{2} polarization feature appears to exhibit two peaks, perhaps associated with the high- and low-velocity components. The position angles across the polarized line features, particularly \ion{Si}{2}, are close to that of the continuum, which suggests an axisymmetric configuration for the SN. 

By day 16.0, the continuum polarization is consistent with having no change relative to day 6.9, while \ion{Si}{2} has increased in strength slightly to peak at this epoch.
A Gaussian fit to the \ion{Si}{2} feature indicates a line polarization of $0.9\pm0.1\%$ with respect to the continuum level. For \ion{Ca}{2} polarization, the enhancement of the high-velocity component from day 6.9 has disappeared and the peak of the lower-velocity component has increased by $\sim0.3$\%.

By day 24.0, the continuum and \ion{Si}{2} line polarization appears to have dropped substantially for wavelengths shortward of 7000\,{\AA}, with no significant change apparent at longer wavelengths; polarization in the continuum region is undetected at this epoch. If real, such a continuum polarization drop roughly one week after peak would be reminiscent of the evolution of SN\,2001el (Wang et al. 2003). However, by day 44.0 the continuum polarization appears to have regained the strength exhibited on day 16.0 and earlier. \ion{Si}{2} has restrengthened as well, while declining in radial velocity along with the minimum of the weakening absorption profile. Based on this unexpected restrengthening, we exercise caution regarding the temporarily weakened polarization on day 24.0, as we are concerned that this could be the result of a systematic error. The drop in polarization appears to have only affected the Stokes $q$ parameter (derived from exposures with polarimeter waveplate angles at 0$^{\circ}$ and 45$^{\circ}$). Each of our three $q$ sequences of the SN are consistent, and we see no such change in the $q$ parameter of our standard-star observations from the same night. Thus, if the change on day 24.0 is the result of systematic error (e.g., some unknown temporary source of instrumental polarization above our typical limit of $<0.05$\%), then it must have occurred over an hourly timescale. Alternatively, a subsequent rise in continuum polarization on day 44.0 could result from the appearance of weak line features in the our chosen continuum region (6700--7150\,{\AA}), but in this case we would not expect the simultaneous rise in the \ion{Si}{2} feature. As a final possibility, the temporary influence of a separate light-echo component, possibly associated with dust in the host ISM, could result in the observed fluctuation; this possibility has the advantage of accounting for the brief change in the continuum and line polarization simultaneously, and it would also explain why the reddest wavelengths are not significantly affected. 

Overall, the spectropolarimetric character of SN\,2016coj is consistent with the trends exhibited by ``normal" SNe~Ia. For example, Maund et al. (2010) reported a correlation between the polarization of the \ion{Si}{2} $\lambda$6355 feature, measured near or before peak luminosity, and the radial-velocity decline rate of the absorption minimum (also see Leonard et al. 2006), physically interpreted as evidence for a single geometric configuration for normal SNe~Ia. At peak brightness on day 16.0, the line polarization of $0.9\pm0.1\%$ combined with our measured value of $-95$\,km\,s$^{-1}$\,day$^{-1}$ for the velocity evolution, shows that SN~2016coj falls where expected on the correlated distribution of SNe~Ia reported by Maund et al. (2010), and within the range of high-velocity explosions.

\begin{table}\begin{center}\begin{minipage}[bp]{3.3in} \setlength{\tabcolsep}{2.8pt}
      \caption{Polarization of SN~2016coj}
\centering
  \begin{tabular}{@{}lcccc@{}} 
  \hline
Epoch     & $P_V\tablenotemark{a}$(\%) & $\theta_V$(deg)  & $P_{\textrm{\tiny{cont}}}$ & $\theta_{\textrm{\tiny{cont}}}$(deg)   \\                            
\hline
\hline
6.9   & 0.27(0.01)& 54.3(0.8)  &  0.51(0.01) & 55.4(0.9) \\
16.0 & 0.20(0.01)& 52.6(0.8)  &  0.38(0.01) & 53.4(0.8) \\
24.0 & 0.16(0.01)& 39.2(1.6)  &  0.06(0.02) & 09.0(4.3) \\
44.0 & 0.39(0.04)& 41.0(1.3)  &  0.28(0.03) & 30.0(2.2) \\
\hline
\end{tabular} \end{minipage} \end{center}
\tablenotetext{a}{$V$-band and continuum integrated over wavelength ranges of 5050--5950\,{\AA} and 6700--7150\,{\AA}, respectively. Uncertainties are statistical.}
\label{hstspitzer_model}
\end{table}

\subsection{High-Resolution Spectra}\label{ss:highresspectra}

\begin{figure*}
\centering
\includegraphics[width=.67\textwidth,angle=270]{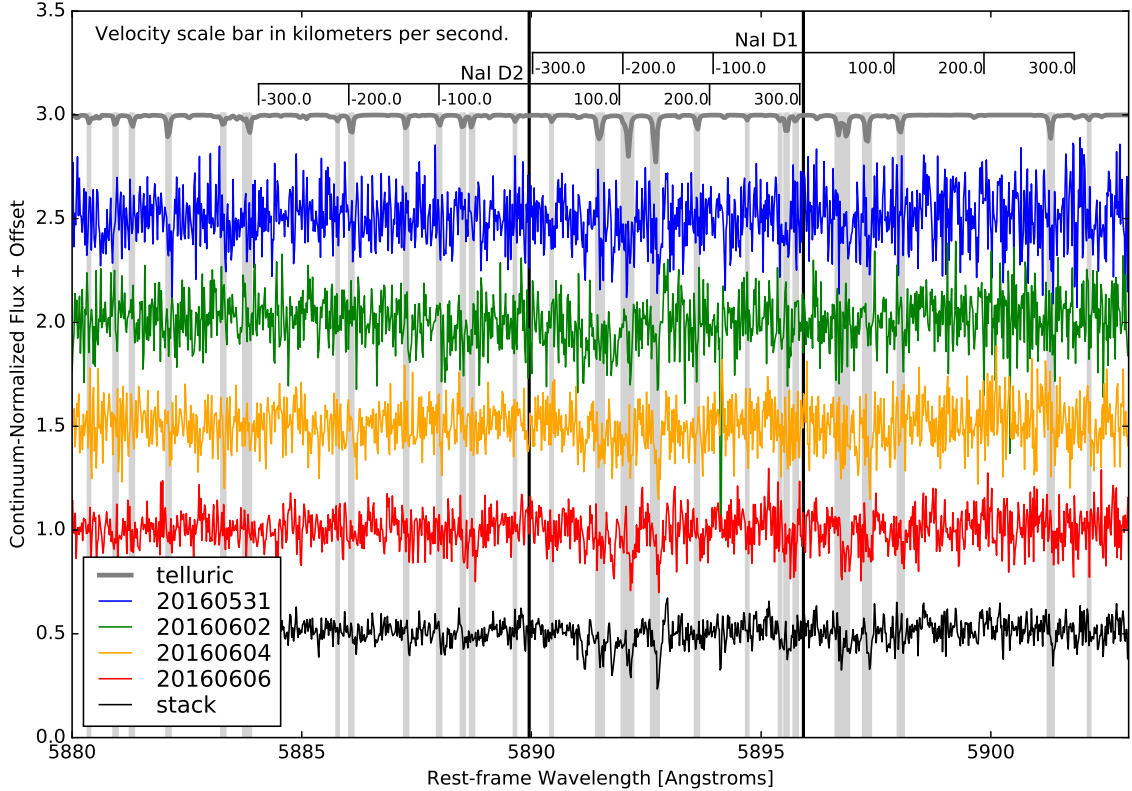}
\caption{High-resolution spectra of SN\,2016coj in the region of \ion{Na}{1}~D absorption from the APF
         obtained at four epochs (2016-05-31, blue; 2016-06-02, green; 2016-06-04, orange; 2016-06-06, red).
	 The bottom (black) shows the stacked spectrum from the data of these epochs.
	 A mean barycentric velocity of $\sim15$\,km\,s$^{-1}$ has been applied
	 for all epochs of SN~2016coj spectra.
	 The atmospheric absorption lines are shown in gray, which has been redshifted
	 into the frame of SN\,2016coj. 
	 The most significant features are highlighted with vertical bars
	 in order to identify their presence in the spectra of SN\,2016coj. A rest-frame velocity scale
	 bar is provided for each \ion{Na}{1}~D feature along the top of the plot.
         \\
         \\
	 }
\label{Fig_spec_high_resolution}
\end{figure*}

We examine the APF high-resolution spectra for narrow absorption features, such as those
that were identified in APF spectra of SN~2014J (Graham et al. 2015). We began spectral
monitoring with the APF based on an early classification and the assumption of a host-galaxy distance smaller than that adopted here.
The object's peak apparent brightness ended up being $\sim3$\,mag fainter than that of SN\,2014J, and
fainter than the projected minimum we typically require for triggering the APF.
For this reason, the S/N of our SN\,2016coj APF spectra is quite low.
Instead of ceasing our APF monitoring we obtained multiple observations over several nights
in order to stack our spectra, but ultimately we do not identify any absorption features
of \ion{Na}{1}~D $\lambda\lambda$5889.95, 5895.92, \ion{Ca}{2}~H\&K $\lambda\lambda$3933.7, 3968.5,
\ion{K}{1} $\lambda\lambda$7664.90, 7698.96, H$\alpha$ $\lambda$6562.801,
H$\beta$ $\lambda$4861.363, or the diffuse interstellar bands
($\lambda \approx 5780$, 5797, 6196, 6283, 6613\,\AA). 

Since the \ion{Na}{1}~D feature
is most useful for constraining the presence of circumstellar material and line-of-sight
host-galaxy dust extinction, and owing to grating blaze is in a region of relatively higher
S/N ($\sim10$), we estimate an upper limit on its $W_\lambda$
in the following way. The flux of the continuum-normalized stacked APF spectra in the region of
\ion{Na}{1}~D, shown in black in Figure \ref{Fig_spec_high_resolution}, has a standard deviation of $\sigma \approx 0.038$.
As an upper limit on the depth of an absorption feature that we could have detected,
we use $3\sigma \approx 0.11$. Our instrumental configuration for the Levy spectrograph results in
a spectral resolution of $\Delta \lambda \approx 0.03$\,\AA, from which we estimate that the minimum
FWHM of a detected feature is $3 \Delta \lambda \approx 0.1$ \AA. Assuming a Gaussian profile
for this hypothetical absorption, we constrain the \ion{Na}{1}~D feature
to have $W_\lambda \lesssim 0.56$\,\AA. Based on Figure 9 of Phillips et al. (2013), this puts an upper limit
on host-galaxy extinction of $A_V\lesssim 0.2$\,mag (with $E(B-V)=0.07$\,mag assuming $R_V=3.1$).
Although this is a rather large upper limit, it is consistent with the small host-galaxy extinction
constrained from our low-resolution spectra (see \S\ref{ss:spectra}) and also with
the low extinction expected given the early type of the host, NGC~4125.


\section{Conclusions}\label{s:conclusions}

In this paper we have presented optical photometric, low- and high-resolution
spectroscopic, and spectropolarimetric observations of SN~2016coj, one
of the youngest discovered and best-observed SNe~Ia.
Our clear-band light curve shows that our first detection is merely
$0.6\pm0.5$\,d after the fitted time of first light, making it one of the earliest detected SNe~Ia.
We estimate that SN~2016coj took $\sim16.0$~d after the fitted first-light time to
reach $B$-band maximum. Its maximum brightness has a normal luminosity,
$B = -18.9\pm0.2$\,mag. An estimated $\Delta m_{15}(B)$ value of 1.25\,mag
along with spectral information support its normal SN~Ia classification.
In the well-observed low-resolution spectral sequence, we identify a
high-velocity feature from both \ion{Ca}{2} H\&K and the \ion{Ca}{2} NIR triplet,
and also possibly from the \ion{O}{1} triplet.
SN~2016coj has a Si~II $\lambda$6355 velocity of $\sim12,600$\,\kms\ at peak 
brightness,
$\sim\,1500$\,\kms\ higher than that of typical SNe~Ia.
We find that the Si~II $\lambda$6355 velocity decreases rapidly
during the first few days and then slowly decreases after peak brightness,
very similar to that of other SNe~Ia.
A broken-power-law function can well fit the Si~II $\lambda$6355 velocity 
for up to about a month after first light.
We estimate there to be very small host-galaxy extinction based on the lack of Na~I~D lines
from the host galaxy in our low- and high-resolution spectra.
Our four epochs of spectropolarimetry show that SN~2016coj exhibits 
weak polarization in the continuum,
but the \ion{Si}{2} line polarization is quite strong ($\sim0.9\pm0.1$\%) at peak brightness.

\begin{acknowledgments}

A.V.F.'s group at U.C. Berkeley is grateful for financial assistance
from National Science Foundation (NSF) 
grant AST-1211916, Gary \& Cynthia Bengier, the Richard \&
Rhoda Goldman Fund, the Christopher R. Redlich Fund, and the TABASGO Foundation.
J.M.S. is supported by an NSF Astronomy and Astrophysics Postdoctoral 
Fellowship under award AST-1302771. 
UC Irvine observing runs were supported in part by NSF grant AST-1412693.
Some of the data presented herein were obtained using the UC Irvine 
Remote Observing Facility, made possible by a generous gift from John and 
Ruth Ann Evans.
V.N.B. and M.C. gratefully acknowledge assistance from an NSF Research
at Undergraduate Institutions (RUI) grant AST-1312296.
The UCSC group is supported in part by NSF grant AST-1518052, and by
fellowships from the Alfred P.\ Sloan Foundation and the David and
Lucile Packard Foundation to R.J.F.
Observations by the UCLA group are supported by NSF grant AST-1412315 to T.T.
X.W. is  financially supported by the National Science Foundation of China
(NSFC grants 11178003 and 11325313). This work was partially supported by
the Open Project Program of the Key Laboratory of Optical Astronomy,
National Astronomical observatories, Chinese Academy of Sciences.
G.H., C.M., and D.A.H. are supported by NSF grant AST-1313484.
We thank the staffs of the various observatories at which data were obtained.
Some of the data presented herein were obtained at the W.~M. Keck 
Observatory, which was made possible by
the generous financial support of the W.~M. Keck Foundation.
Based in part on observations obtained at Kitt Peak National Observatory,
National Optical Astronomy Observatory (NOAO Prop. ID 2015B-0313; PI Foley),
which is operated by the Association of Universities for Research in 
Astronomy (AURA), Inc., under cooperative agreement with the NSF.
The authors are honored to be permitted to conduct astronomical research on 
Iolkam Du'ag (Kitt Peak), a mountain with particular significance to 
the Tohono O'odham.
This work makes use of observations from the LCOGT network.
We thank Lauren Weiss and Bradford Holden for obtaining APF data at
Lick Observatory.
Research at Lick Observatory is partially supported by a generous gift from Google.

\end{acknowledgments}

\end{document}